\documentclass[reprint,superscriptaddress]{revtex4-1}

\usepackage{graphicx}
\usepackage{amsmath}

\begin{document}

\title{Extending the frequency range of digital noise measurements to the microwave domain}

\author{Stephen R. Parker}
\email{stephen.parker@uwa.edu.au}
\affiliation{School of Physics, The University of Western Australia, Crawley 6009, Australia}
\author{Eugene N. Ivanov}
\affiliation{School of Physics, The University of Western Australia, Crawley 6009, Australia}
\author{John G. Hartnett}
\affiliation{School of Physics, The University of Western Australia, Crawley 6009, Australia}
\affiliation{Institute of Photonics and Advanced Sensing, School of Chemistry and Physics, University of Adelaide, Adelaide 5005, Australia}


\begin{abstract}
We describe the use of digital phase noise test sets at frequencies well beyond the sampling rate of their analog-to-digital converters. The technique proposed involves the transfer of phase fluctuations from an arbitrary high carrier frequency to within the operating frequency range of the digital instrument. The validity of the proposed technique has been proven via comparison with conventional methods. Digital noise measurements eliminate the need for calibration and improve consistency of experimental results. Mechanisms limiting the resolution of spectral measurements are also discussed.
\end{abstract}

\maketitle

\section{Introduction}

Over the past two decades the technique of noise measurements has been refined to the point where its spectral resolution exceeded the standard thermal noise limit~\cite{Realtime}. Yet in many practical cases the extreme spectral resolution is not required. Besides, high-resolution phase noise measurements are time consuming and technically involved; they also suffer from the residual sensitivity to amplitude fluctuations of the carrier signal. 

Recent years have seen the introduction of digital test sets that allow fast and reasonably accurate measurements of differential phase fluctuations between two input signals~\cite{Stein2004,Stein2010}. This is achieved via the coherent demodulation of the input signals. During this process, an input signal u($t$)~=~U~Cos$\left(\omega_{\text{sign}}t+\Phi(t)\right)$ is sampled and its frequency is measured. The digital replica of the input signal u($t_{\text{k}}$) is then multiplied with a synthesized discrete cosine and sine wave at frequency $\omega_{\text{sign}}$. The following low-pass filtering eliminates the high frequency spectral components (at 2~$\omega_{\text{sign}}$) leaving only slowly varying terms x($t_{\text{k}}$)~=~Sin$\left(\Phi(t_{\text{k}})\right)$ and y($t_{\text{k}}$)~=~Cos$\left(\Phi(t_{\text{k}})\right)$. The phase of the input signal is then computed via the inverse trigonometric function: $\Phi(t_{\text{k}})$~=~ArcTan$\left(\text{x}(t_{\text{k}})/\text{y}(t_{\text{k}})\right)$. To reduce the excess phase noise associated with the sampling process, digital test sets feature multi-channel architecture complemented by the cross-correlation signal processing~\cite{WallsCC,Rubiola}.

When applied to the characterization of phase noise in oscillators, digital test sets simplify the measurement process by eliminating the need for phase-locking one oscillator to another~\cite{Hartnett2013}. Among their advantages is also immunity to amplitude fluctuations and the ability to self-calibrate. The main drawback of digital test sets is the restricted frequency range: the current commercial instruments operate below 400~MHz~\cite{Symmetricom}. 

In this paper we describe a technique that enables high-resolution phase noise measurements with digital instruments at frequencies well beyond the sampling rate of their analog-to-digital converters. Microwave amplifiers are used as test objects to verify the validity of these digital noise measurements. We show that the results of digital noise measurements are very much consistent with those obtained using the conventional analog based instruments. We also discuss mechanisms limiting the spectral resolution of noise measurements involving the use of the digital phase noise test sets.

\section{Measurement technique} 

\begin{figure}[t!]
\centering
\includegraphics[width=0.95\columnwidth]{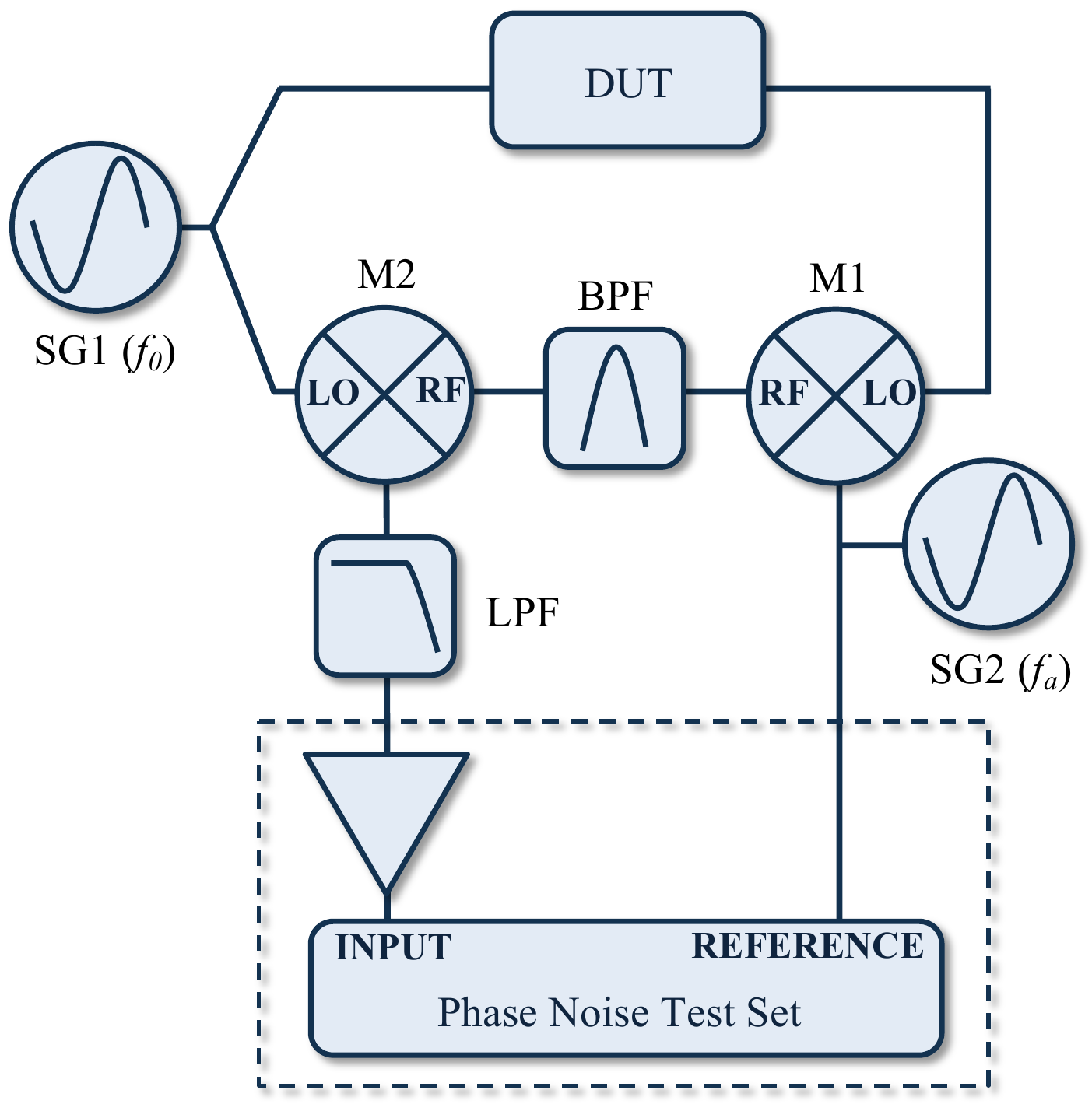}
\caption{Schematic of the transposed frequency circuit (isolators and attenuators not shown). The dashed line rectangle shows the addition of a low noise amplifier and phase noise test set as discussed in the text.}
\label{fig:tf}
\end{figure}

A schematic diagram of the proposed noise measurement system is shown in Fig.~\ref{fig:tf}. Noise fluctuations from a two port Device Under Test (DUT) are introduced to a high frequency carrier (SG1), which is then mixed (M1) with a lower frequency offset (SG2) to  create two sidebands. One sideband is removed via filtering (BPF) before the signal is mixed (M2) with the original high frequency carrier (SG1) and filtered again (LPF), leaving just the low frequency signal and the noise fluctuations of the DUT. In essence, the circuit transfers the noise components from one frequency to a different frequency, as such we refer to the setup as a transposed frequency noise measurement system.

In the analysis that follows we represent waveforms as cosine functions, having dropped the sine components to allow for a clear and concise display for the reader. In addition, most amplitude coefficients have been set to unity.

The signal of the high frequency oscillator (SG1) is modelled as
\begin{equation}
\cos{\left(\omega_{0}t+\delta\phi_{\text{SG1}}\right)},
\label{eq:SG1}
\end{equation}
where $\omega_{0}=2\pi f_{0}$ represents the oscillator frequency and $\delta\phi_{\text{SG1}}$ are the intrinsic phase fluctuations of the signal generator. Eq.~\eqref{eq:SG1} is then power divided into two paths, with one arm passing through the DUT where phase fluctuations of the device $\delta\phi_{\text{DUT}}$ are imprinted on to the carrier,
\begin{equation}
\cos{\left(\omega_{0}t+\delta\phi_{\text{SG1}}+\delta\phi_{\text{DUT}}+\phi_{\text{DUT}}\right)},
\label{eq:DUT}
\end{equation}
where $\phi_{\text{DUT}}$ is the mean (fixed) phase delay in the DUT. The output of the DUT is then mixed (M1) with an auxiliary carrier from the signal generator SG2 at a lower frequency $\omega_{a}=2\pi f_a$ that falls within the bandwidth of the digital phase noise test set,
\begin{align}
\cos{\left(\omega_{0}t+\delta\phi_{\text{SG1}}+\delta\phi_{\text{DUT}}+\phi_{\text{DUT}}\right)}\times\cos{\left(\omega_{a}t+\delta\phi_{\text{SG2}}\right)} \nonumber \\
=\cos{\left(\left(\omega_{0}+\omega_{a}\right)t+\delta\phi_{\text{SG1}}+\delta\phi_{\text{DUT}}+\phi_{\text{DUT}}\right.} \nonumber \\
\left.+\delta\phi_{\text{SG2}}+\delta\phi_{\text{M1}}\right) \label{eq:sideband2} \\
+\cos{\left(\left(\omega_{0}-\omega_{a}\right)t+\delta\phi_{\text{SG1}}+\delta\phi_{\text{DUT}}+\phi_{\text{DUT}}\right.} \nonumber \\
\left.-\delta\phi_{\text{SG2}}+\delta\phi_{\text{M1}}\right), \nonumber
\end{align}
with $\delta\phi_{\text{M1}}$ being the phase noise contribution of mixer M1. Assuming that a band-pass filter (BPF) eliminates the upper sideband of the beat note at the output of mixer M1, the signal at the output of mixer M2 can be written as
\begin{align}
\cos{\left(\left(\omega_{0}-\omega_{a}\right)t+\delta\phi_{\text{SG1}}+\delta\phi_{\text{DUT}}+\phi_{\text{DUT}}\right.} \nonumber \\
\left.-\delta\phi_{\text{SG2}}+\delta\phi_{\text{M1}}\right)\times\cos{\left(\omega_{0}t+\delta\phi_{\text{SG1}}\right)} \nonumber \\
=\cos{\left(\left(2\omega_{0}-\omega_{a}\right)t+2\delta\phi_{\text{SG1}}+\delta\phi_{\text{DUT}}+\phi_{\text{DUT}}\right.} \label{eq:m2sb2} \\
\left.-\delta\phi_{\text{SG2}}+\delta\phi_{\text{M1}}+\delta\phi_{\text{M2}}\right) \label{eq:m2sb1} \nonumber \\
+\cos{\left(\omega_{a}t+\delta\phi_{\text{DUT}}+\phi_{\text{DUT}}+\delta\phi_{\text{SG2}}+\delta\phi_{\text{M1}}+\delta\phi_{\text{M2}}\right)}. \nonumber
\end{align}
The low-pass filter (LPF) in Fig.~\ref{fig:tf} allows only a signal at frequency $f_a$ to reach the input of the digital test set,
\begin{equation}
\cos{\left(\omega_{a}t+\delta\phi_{\text{DUT}}+\phi_{\text{DUT}}+\delta\phi_{\text{SG2}}+\delta\phi_{\text{M1}}+\delta\phi_{\text{M2}}\right)}.
\label{eq:lpfout}
\end{equation}
\begin{figure}[t!]
\centering
\includegraphics[width=0.85\columnwidth]{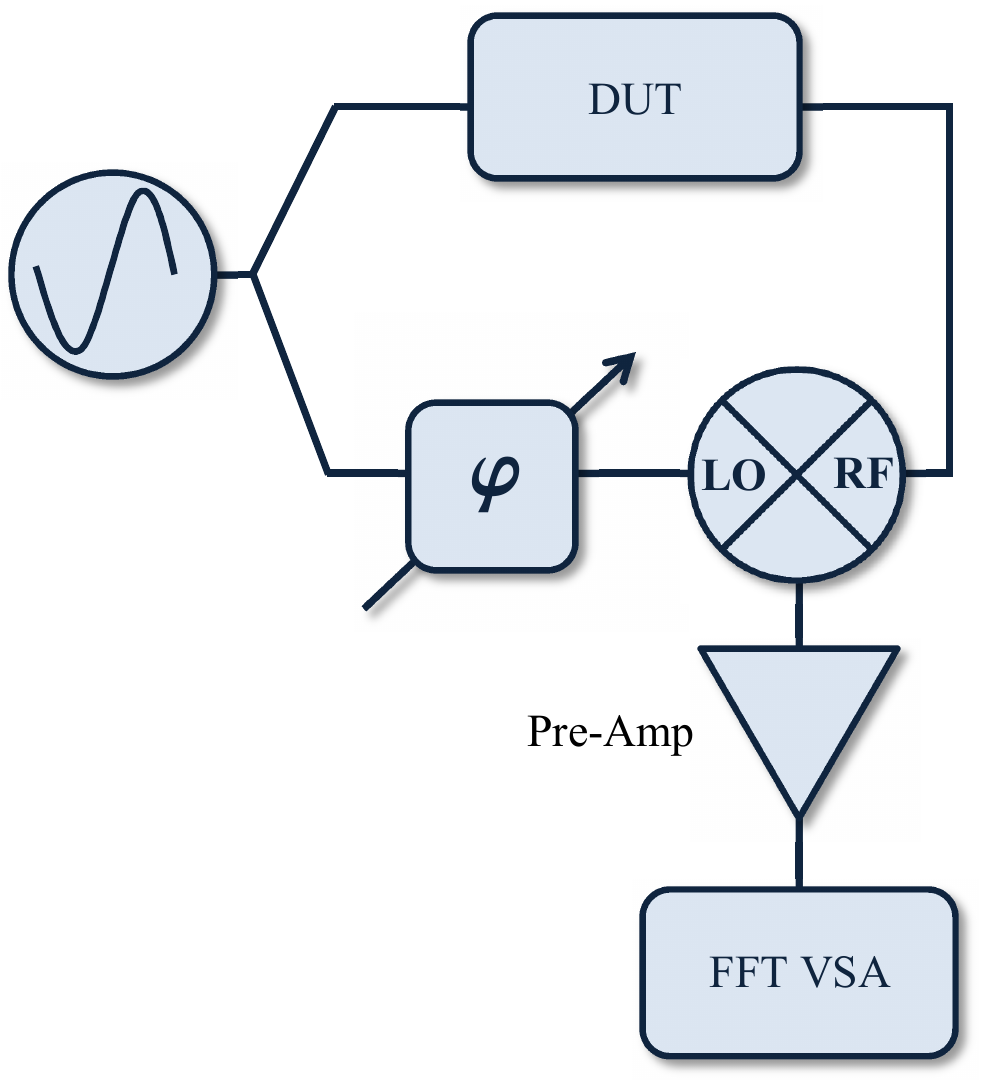}
\caption{Schematic of a phase bridge with a low noise pre-amplifier (Stanford Research Systems SR560) and Fast Fourier Transform Vector Signal Analyzer (FFT VSA, Agilent 89410A). The mechanical phase shifter can be adjusted to make the mixer output proportional to phase or amplitude fluctuations in the DUT.}
\label{fig:FFT}
\end{figure}
This signal contains all information about the phase fluctuations of the microwave DUT and is free from phase fluctuations of the microwave pump source SG1. The phase fluctuations of the auxiliary source SG2 are not an issue either: they are synchronous at both ports of the test set and cancel each other upon subtraction. On the other hand, phase fluctuations of the two mixers, as follows from~\eqref{eq:lpfout}, are in direct competition with the useful phase noise of the DUT. For this reason, one could assume that mixer noise would be the main factor limiting the resolution of spectral measurements (this is confirmed by experimental results of section~\ref{sec:dis}). 
In deriving~\eqref{eq:m2sb2} it was assumed that phase noise of the microwave pump oscillator is canceled at the output of mixer M2. In reality, such cancellation may not be perfect due to asymmetry of the measurement system and associated with it differential time delay $\tau$ between the microwave signals ($\omega_{0}$) at the LO and RF ports of the mixer M2. In this scenario, the Power Spectral Density (PSD) of the residual phase fluctuations of the microwave oscillator can be estimated from
\begin{equation}
S^{res}_{\phi}\left(F\right)\sim S^{\mu W}_{\phi}\left(F\right)\left(\pi F \tau\right)^{2},
\label{eq:osc}
\end{equation}
where S$^{\mu W}_{\phi}\left(F\right)$ is the PSD of phase fluctuations of the microwave oscillator and $F$ is the Fourier frequency. Eq.~\eqref{eq:osc} is valid for phase fluctuations with coherence time significantly longer than $\tau$. This, however, is always true in the microwave domain where the oscillator frequency experiences slow modulation, such as flicker or random walk fluctuations. Considering a quartz-based frequency synthesizer, such as an Agilent E8257D operating at 10~GHz, the PSD of its phase fluctuations is -70~dBc/Hz at F=10~Hz~\cite{Agilent2}. Further assuming that $\tau$~=~10~ns (corresponding to an excessively large path length difference of 3~m) and making use of~\eqref{eq:osc} we find that the PSD of the residual phase fluctuations is totally negligible (less than -190~dBc/Hz at F~=~10~Hz), as compared to the mixer contribution (-135...-130~dBc/Hz at F~=~10~Hz).

In the case when the power of the output signal at frequency $f_a$ is not sufficient to satisfy the power requirements of the phase noise test set, it can be boosted with a low-noise RF amplifier (see Fig.~\ref{fig:tf}). This could markedly limit the resolution of spectral measurements at relatively high Fourier frequencies (F~$>$~1~kHz) due to the amplifier Nyquist noise (see the noise spectra in Fig.~\ref{fig:tfresults}).

If both sidebands at $f_{0}\pm f_{a}$ (Eq.~\eqref{eq:sideband2}) are present for the second mixing stage M2, i.e. if no BPF is used, then the phase fluctuations of the DUT are converted in to amplitude fluctuations of the output signal at $f_{a}$,
\begin{align}
\cos{\left(\left(\omega_{0}\pm \omega_{a}\right)t+\delta\phi_{\text{SG1}}+\delta\phi_{\text{DUT}}+\phi_{\text{DUT}}\right.} \nonumber \\
\left.+\delta\phi_{\text{SG2}}+\delta\phi_{\text{M1}}\right)\times\cos{\left(\omega_{0}t+\delta\phi_{\text{SG1}}\right)} \nonumber \\
=2\cos{\left(\delta\phi_{\text{DUT}}+\phi_{\text{DUT}}+\delta\phi_{\text{SG2}}+\delta\phi_{\text{M1}}+\delta\phi_{\text{M2}}\right)} \nonumber \\
\times\cos{\left(\omega_{a}t\right)}. \label{eq:am}
\end{align}
As follows from~\eqref{eq:am}, the phase of the output signal is independent of the phase of the DUT, therefore, all information about the phase fluctuations of the DUT is lost. In this regime one can evaluate the noise floor of the measurement system. On the other hand, the dependence of the amplitude of the output signal on the phase of either the DUT or that of the  auxiliary signal generator SG2 could be used for generation of pure AM-modulated signals.

\section{Results}

\begin{figure}[t!]
\centering
\includegraphics[width=0.95\columnwidth]{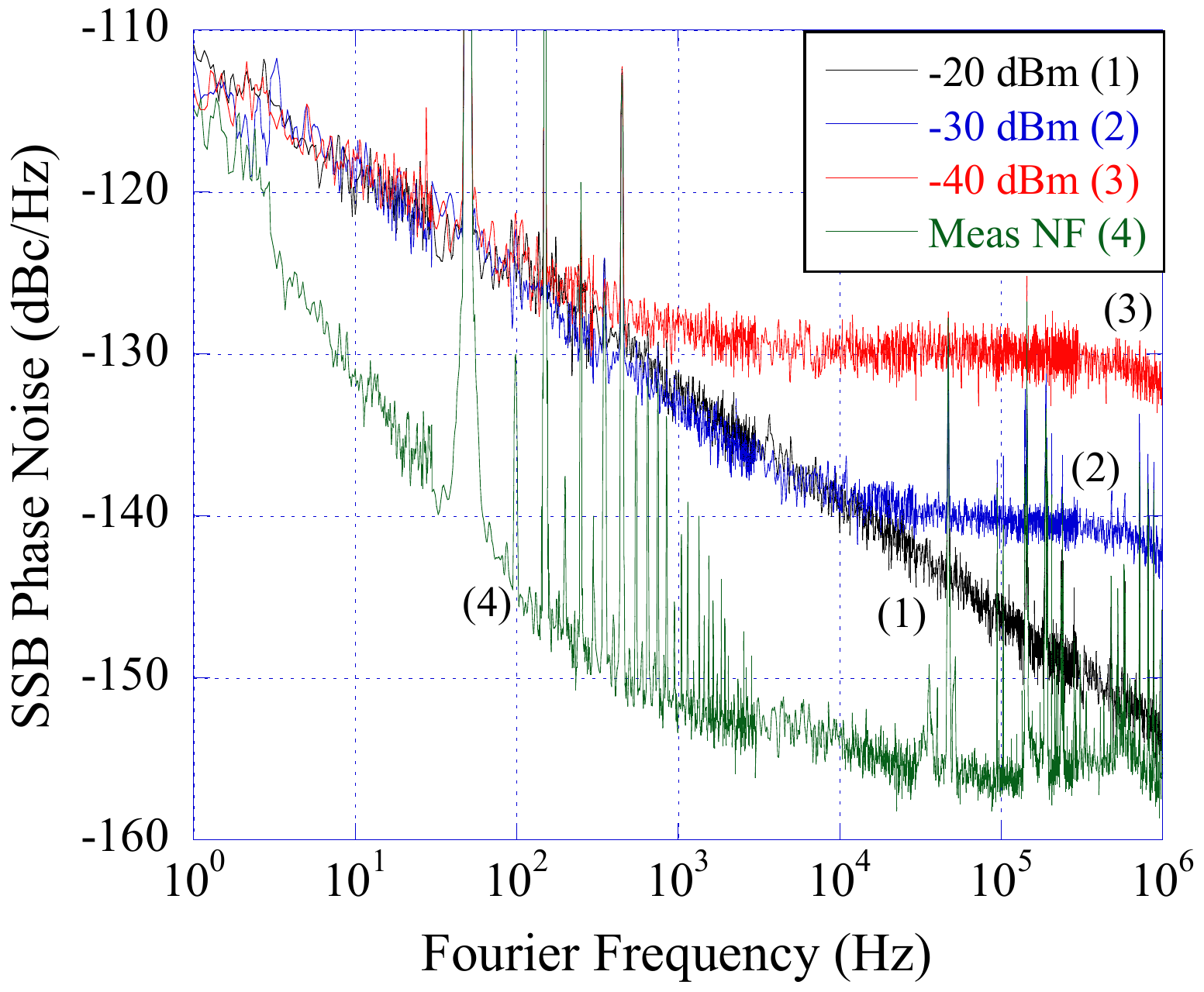}
\caption{(color online) Single-sideband phase noise of an Endwave JCA812-5001 amplifier with different input power (curves 1 -  3, see legend), measured using a phase bridge system (Fig.~\ref{fig:FFT}) with a carrier signal frequency of 11.2~GHz. Curve 4 shows the measurement system noise floor. }
\label{fig:FFTresults}
\end{figure}

In order to verify the validity of the transposed frequency noise measurement technique we applied it to the case of a microwave amplifier with well known noise properties.

Amplifier noise properties are often characterized by a Noise Figure (NF) in which case the Single Sideband (SSB) PSD of amplifier white phase noise is given by~\cite{dicke1946,ivanov1996}
\begin{equation}
\mathcal{L}_{\phi}^{W} = \frac{k_{B}T_{0}}{2\text{P}_{\text{in}}}\times\text{NF},
\label{eq:ampthermal}
\end{equation}
where $k_{B}$ is Boltzmann's constant, $T_{0}$ is the ambient temperature of the amplifier, $P_{\text{in}}$ is the input power and NF is the Noise Factor, which is the Noise Figure represented as a multiplicative factor instead of in decibel units. Amplifiers also exhibit flicker (1/$F$) phase noise, so for a complete understanding of amplifier noise performance one must determine the corner frequency where the phase noise transitions from a flicker frequency dependence to a white frequency dependence. The amplifier to be measured is a JCA812-5001, originally manufactured by Endwave Defense Systems (now owned by Microsemi Corporation). The amplifier operates over a frequency range of 8 - 12 GHz with a gain of 48 dB, a compression point power of 14.5 dBm and a noise figure of 5 dB. From eq.~\eqref{eq:ampthermal} we expect the amplifier to generate a SSB white phase noise floor of -172 - P$_{\text{in}}$ dBc/Hz, where P$_{\text{in}}$ is expressed in units of dBm.

\begin{figure}[t!]
\centering
\includegraphics[width=0.95\columnwidth]{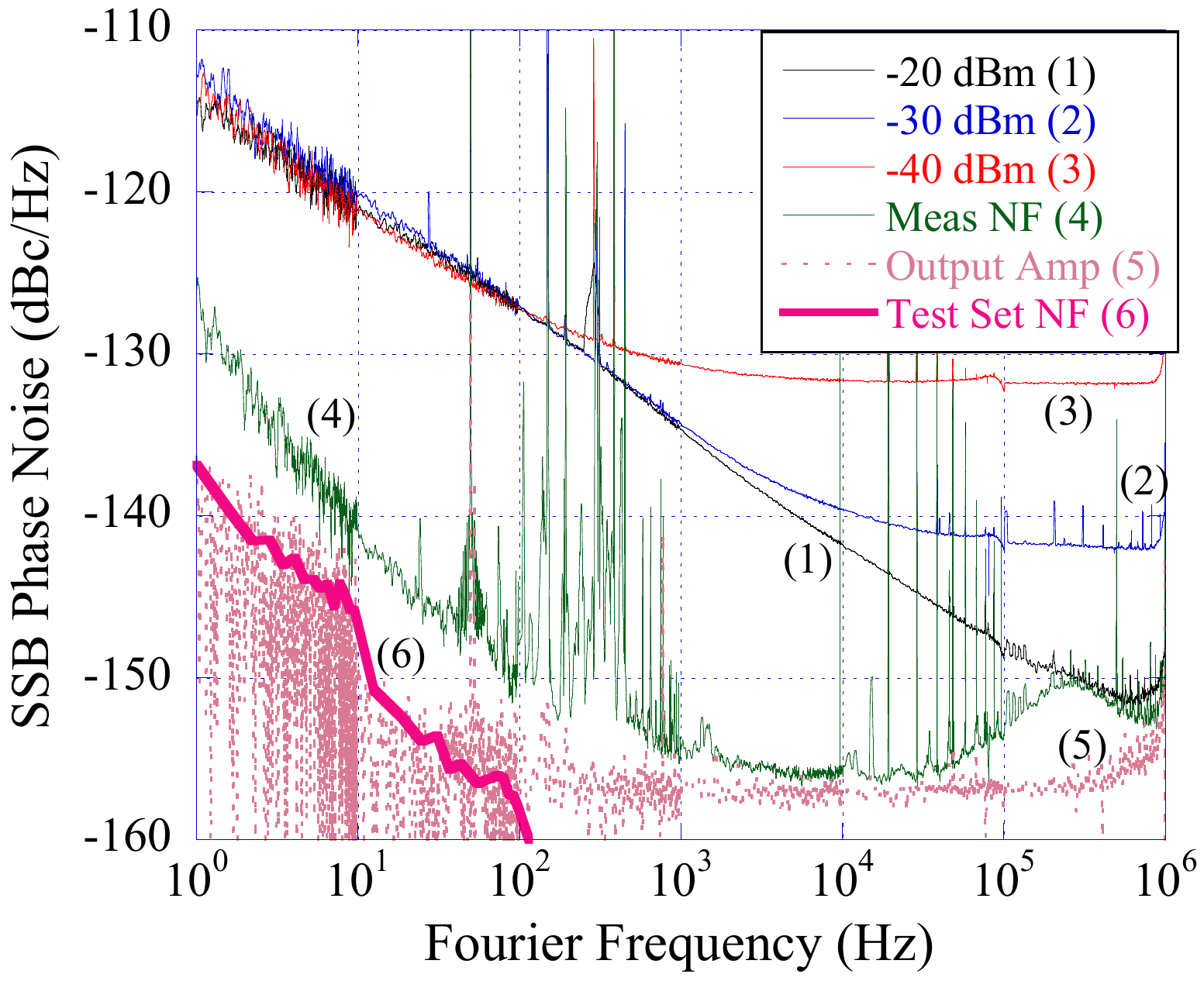}
\caption{(color online)  Single-sideband phase noise of an Endwave JCA812-5001 amplifier for different input powers (curves 1 -  3, see legend), measured using a digital phase noise test set~\cite{Symmetricom} combined with a transposed frequency system (Fig.~\ref{fig:tf}) with a carrier signal frequency of 11.2 GHz and an auxiliary signal frequency of 20 MHz. Curve 4 shows the total measurement system noise floor, curve 5 is the phase noise of the final output amplifier and curve 6 is the noise floor of the test set.}
\label{fig:tfresults}
\end{figure}

First, a phase bridge measurement system (Fig.~\ref{fig:FFT}) was used to measure the phase noise of the amplifier. The mechanical phase shifter is adjusted so that the two signals incident on the mixer are in quadrature. In this regime, the mixer output DC voltage is close to zero, while its fluctuations are synchronous with the phase fluctuations of the DUT. This output passes through a low noise pre-amplifier before being collected by a Fast Fourier Transform Vector Signal Analyzer (FFT VSA). The SSB PSD of voltage fluctuations, $S_V(F)$, can then be converted in to SSB phase noise by
\begin{equation}
\mathcal{L}_{\phi}\left(F\right) = S_{V}\left(F\right) - S_{pd} - \kappa_{\text{PA}}-3,
\label{eq:pnFFT}
\end{equation}
where $\kappa_{\text{PA}}$ is the gain of the pre-amplifier and $S_{pd}$ is the phase-to-voltage conversion efficiency of the phase bridge, both converted into dB units. The value of $S_{pd}$ depends upon the choice of mixer and the power levels incident on the mixer; it can be measured by adjusting the mechanical phase shifter and observing the change in voltage at the output of the mixer. For the phase bridge used in these measurements $S_{pd}\approx$~0.22~V/rad.
In-line attenuators were used to maintain a constant power level at the RF port of the mixer in order to avoid repeating the calibration of the phase bridge. The noise floor of the combined phase bridge/pre-amplifier/VSA system was measured by removing the DUT and replacing it with coaxial cable. Phase noise of the amplifier operating at 11.2~GHz was measured for different input powers, the results are presented in Fig.~\ref{fig:FFTresults}. None of the results above 1~Hz offset are limited by the noise floor of the measurement system. The rolloff observed above 10$^{5}$~Hz is due to filtering in the pre-amplifier. The observed levels of white phase noise are congruent with the predictions of eq.~\eqref{eq:ampthermal}.

The transposed frequency circuit in Fig.~\ref{fig:tf} was used in conjunction with a digital phase noise test set~\cite{Symmetricom} to repeat the measurements of Fig.~\ref{fig:FFTresults}. The test set used had an input bandwidth of 1 - 400~MHz and required a minimum signal power of 3~dBm. Conventional microwave mixers are not capable of satisfying the test set power requirements, as such a low noise amplifier was used to increase the output power. Signal generator SG1 operated at 11.2~GHz and  signal generator SG2 operated at 20~MHz, which also acted as a reference signal for the test set. The bandpass filter was adjusted to preserve the 11.18~GHz sideband. In-line attenuators were used to ensure that power incident on all mixer ports remained constant for all measurements. It should be noted that passive two port microwave devices such as in-line attenuators and isolators generally exhibit extremely low levels of residual phase noise~\cite{inter} and hence make a negligible contribution to the noise floors of the measurement systems studied here. Phase noise measurements of the amplifier obtained with the measurement setup in Fig.~\ref{fig:tf} are presented in Fig.~\ref{fig:tfresults}, the only result limited by the noise floor of the system is for an input power of -20~dBm at offset frequencies above 10$^{5}$~Hz.

\section{Discussion} \label{sec:dis}

A comparison of Fig.~\ref{fig:FFTresults} and Fig.~\ref{fig:tfresults} does not indicate any discrepancies. Differences between the white noise floors can be attributed to minor inconsistencies in the power incident on the DUT. The data presented in Fig.~\ref{fig:FFTresults} was only averaged from 16 - 64 times depending on the frequency offset while the data presented in Fig.~\ref{fig:tfresults} was averaged on the order of 100 - 400 averages depending on the frequency offset. This does not represent any fundamental disadvantage of either approach but is simply a reflection of differences in acquisition time. Regardless, the results demonstrate that the transposed frequency technique correctly measures the phase noise of the DUT. It does not matter if the DUT is located in the other arm from the power divider as the same result is achieved.
Curve 4 of Fig.~\ref{fig:tfresults} shows the noise floor of the measurement system in Fig.~\ref{fig:tf}. These measurements were made with the DUT replaced by a piece of coaxial cable. It is clear that the intrinsic noise of the test set (curve 6) does not limit the spectral resolution of the overall measurement system. 
At Fourier frequencies above 1~kHz the SSB phase noise floor is flat and close to -157~dBc/Hz. This limit comes from the intrinsic fluctuations of the RF amplifier: these fluctuations were measured in separate experiments at powers of the input signal corresponding to the typical operating conditions. 
At Fourier frequencies below 10$^{3}$~Hz, the phase noise floor of the measurement system is entirely determined by voltage fluctuations in the mixers M1 and M2. This follows from the direct comparison of noise floors of the digital (Fig.~\ref{fig:tf}) and analog (Fig.~\ref{fig:FFT}) measurement systems. Furthermore, curve 4 in Fig.~\ref{fig:FFTresults} indicates that the phase bridge was visibly disturbed during the noise floor measurements. Typically, the intrinsic SSB phase noise of the X-band phase bridge varies as $4\times10^{-13}/F$~(rad$^{2}$/Hz) resulting in $\mathcal{L}$(1~Hz)$\sim-125$~dBc/Hz, which is more than 10~dB lower than what was measured.
The excess noise at frequencies 10$^{5}$~-~10$^{6}$~Hz (curve 4 in Fig.~\ref{fig:tfresults}) could be attributed to either of the signal generators. Making the measurement system more compact should rectify this problem. 
Finally, a consideration must be given to the choice of the auxiliary frequency, $f_a$, as the two sidebands generated after the first mixing stage must be sufficiently separated in order for the band-pass filter to fully attenuate one of them. Assuming that high-quality band-pass filters based on iris-coupled cavity resonators are used, the narrowest bandwidth achievable at the X-band is of the order 50~MHz. So, as long as the auxiliary frequency is larger than 25~MHz one does not have to worry about the loss of sensitivity while making the phase noise measurements. 

\section{Conclusion}

We have developed a novel noise measurement technique, which enables the transfer of phase information from a very high carrier frequency to a much lower one at which this information can be analyzed with digital instruments. The spectral resolution of the entirely digital noise measurement system was found to be similar to that of a conventional phase bridge. Yet, among the advantages of the digital measurements is the immunity to intensity noise and the ability of self-calibration, which streamlines the noise measurement process.

\begin{acknowledgments}
The authors thank Jean-Michel LeFloch for the data acquisition software. This work was supported by Australian Research Council grants LP110200142 and DP130100205.  
\end{acknowledgments}

\end{document}